\documentclass{article}

\usepackage[final]{neurips_ai4d3_2023}
\usepackage{multirow}

\usepackage[utf8]{inputenc} 
\usepackage[T1]{fontenc}    
\usepackage{hyperref}       
\usepackage{url}            
\usepackage{booktabs}       
\usepackage{amsfonts}       
\usepackage{amsmath}
\usepackage{nicefrac}       
\usepackage{microtype}      
\usepackage{xcolor}         
\usepackage{graphicx} 
\usepackage{placeins}
\usepackage{svg}

\newcommand{\igdiff}{\textit{IgDiff}}
\newcommand{\framediff}{\textit{FrameDiff}}
\newcommand{\rfdiff}{RFDiffusion}
\newcommand{\abbtwo}{ABodyBuilder2}
\newcommand{\abbtwoshort}{ABB2}

\title{De novo antibody design with SE(3) diffusion}

\author{
Daniel Cutting$^{1*}$ \quad Frédéric A. Dreyer$^{1*}$ \quad David Errington$^1$ \quad Constantin Schneider$^1$ \\
\textbf{Charlotte M. Deane}$^{1,2}$ \\
$^1$Exscientia, Oxford Science Park, Oxford, OX4 4GE, UK \\
$^2$Department of Statistics, University of Oxford, Oxford OX1 3LB, UK\\
\texttt{\{dcutting,fdreyer,derrington,cschneider\}@exscientia.co.uk}\\
\texttt{deane@stats.ox.ac.uk}
}

\begin{document}

\maketitle

\def\thefootnote{*}\footnotetext{Equal contribution.}
\begin{abstract}
We introduce {\igdiff}, an antibody variable domain diffusion model based on a general protein backbone diffusion framework which was extended to handle multiple chains.
Assessing the designability and novelty of the structures generated with our model, we find that {\igdiff} produces highly designable antibodies that can contain novel binding regions.
The backbone dihedral angles of sampled structures show good agreement with a reference antibody distribution.
We verify these designed antibodies experimentally and find that all express with high yield.
Finally, we compare our model with a state-of-the-art generative backbone diffusion model on a range of antibody design tasks, such as the design of the complementarity determining regions or the pairing of a light chain to an existing heavy chain, and show improved properties and designability.
\end{abstract}

\section{Introduction}
Engineering novel proteins that can satisfy specified functional properties is the central aim of rational protein design.
While sequence-based methods have seen some success~\citep{sequencegen}, they are intrinsically limited by the fact that most properties of a molecule, such as binding or solubility, are determined by their three-dimensional structure.
Recent advances in diffusion models~\citep{ddpm,dsm}, a class of deep probabilistic generative models, have shown promise as a data-driven alternative to more computationally expensive physics-based methods~\citep{rosetta} in tackling \textit{de novo} protein design.
Most approaches focus on modelling only the backbone~\citep{rfdiffusion,genie}, while the sequence is inferred through an inverse folding model, though some full-atom models have been explored~\citep{protpardelle,abdiffuser}.

An application of particular therapeutic relevance is the design of immunoglobulin proteins, which play a central role in helping the adaptive immune system identify and neutralise pathogens. 
They consist of two heavy and two light chains. 
These are separated into constant domains that specify effector function, and a variable domain that contains six hypervariable loops, known as the complementarity determining regions (CDRs), which control binding specificity.
The third loop of the heavy chain, CDR-H3, is the most structurally diverse domain of the antibody, and often determines antigen recognition~\citep{cdrh3_2011,cdrh3}. 
Monoclonal antibodies are an emerging drug modality with the potential for applications in a wide range of therapeutic areas, for example oncogenic, infectious and autoimmune diseases.
They can be adapted to target specific antigens or receptors through engineering of the binding site~\citep{engreview}.

In this article, we consider the recent backbone diffusion model {\framediff}~\citep{framediff} and adapt it for applications to antibody variable regions. We then fine-tune it on synthetic antibody structures from the ImmuneBuilder dataset~\citep{abb2}.
Our {\igdiff} model is trained to generate the paired variable region backbone, with a paired heavy and light chain. 
We study the designability and novelty of the structures generated by our heavy chain model and predict the corresponding sequences with AbMPNN~\citep{abmpnn}, an antibody-specific inverse folding model based on ProteinMPNN~\citep{proteinmpnn}.
We then consider a range of design tasks and show how {\igdiff} can outperform existing backbone diffusion methods on antibody engineering tasks.

\section{Related work}
Denoising diffusion probabilistic models~\citep{diffusion,ddpm} have shown promising results to the task of \textit{de novo} protein design.
Recent models based on a $C_\alpha$-only backbone representation of proteins~\citep{smcdiff} have already been adapted to account for the full frame representation of backbone residues~\citep{framediff, genie} through the application of diffusion on Riemannian manifolds~\citep{riemannian}.
Notably, tools such as {\rfdiff}~\citep{rfdiffusion}, which rely on a fine-tuned RoseTTAFold structure prediction network for the reverse diffusion process, can tackle a number of complex protein engineering tasks such as protein binder design, enzyme active site scaffolding or topology-constrained protein monomer design.
These diffusion models often operate in structural space, and an inverse folding model such as ProteinMPNN~\citep{structransfo,proteinmpnn} is needed to recover sequences corresponding to the predicted structures.
To reconstruct the placement of side-chain atoms, physics-based methods packing tools such as Rosetta~\citep{rosetta} can be used.
AI-driven side-chain packing methods based on SE(3) transformers~\citep{se3transformers}, or a diffusion model on the joint distribution of torsional angles~\citep{diffpack} can also offer an efficient alternative.

Alternatively, protein diffusion models have been applied in sequence space to directly generate amino acid sequences~\citep{proteindesign} which can allow for sequence specific attributes and functional properties.
Classifier-guided discrete diffusion models such as NOS~\citep{lambo2} can for example leverage a discriminative model and the large amount of available sequence data to predict sequences with high fitness.

Antibody design has seen rapid progress and will benefit from these advances in generative protein models.
This task was traditionally tackled using energy-based optimization strategies~\citep{10.1371/journal.pone.0105954,10.1371/journal.pcbi.1006112}, and more recently with sequence-based language models~\citep{10.1093/bioinformatics/btz895,lstmab}.
Several structure-based approaches relying on graph neural networks, such as RefineGNN~\citep{refinegnn}, HSRN~\citep{hsrn}, MEAN~\citep{mean} and its subsequent end-to-end design model dyMEAN~\citep{dymean}, or sequence and structure co-design diffusion models~\cite{diffab}, have shown promising results in predicting the CDR design of antibodies to target a specific epitope on an antigen.
The {\framediff} framework has also been recently explored for inpainting CDR loops of T-cell receptors~\citep{framedipt}, while other recent work has explored the application of structure-based diffusion models to the design of nanobodies~\citep{bakerAb} and antibodies~\citep{abdiffuser}.

\section{SE(3) protein backbone diffusion model}

We review the SE(3) diffusion framework introduced by~\citet{framediff}, which constructs an explicit framework for the diffusion of protein backbones based on the Riemannian score-based generative modeling approach of~\citet{riemannian}.

\begin{figure}
    \centering
    \includegraphics[width=0.9\linewidth]{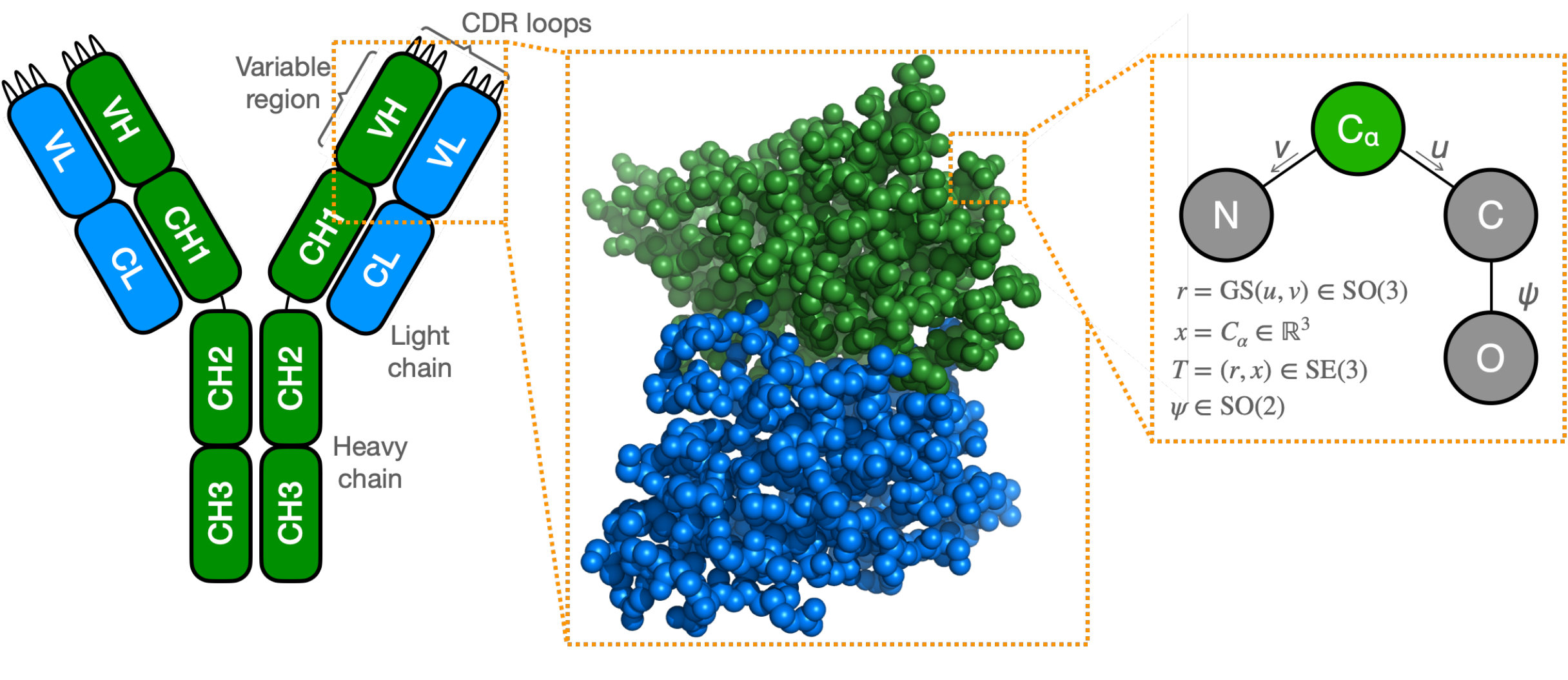}
    \caption{Left: Schematic representation of an antibody. Centre: Backbone atoms of the variable region, showing both a heavy (green) and light (blue) chain domain. Right: The parametrisation of residues into frames used by the diffusion model, with each frame consisting of four heavy atoms connected by rigid covalent bonds.}
    \label{fig:representation}
\end{figure}

For the backbone frame parametrisation we adopt the same formalism as in AlphaFold2~\citep{af2}, using a collection of $N$ orientation preserving rigid transformations to represent an $N$ residue backbone, as shown in figure~\ref{fig:representation}.
These frames map from fixed coordinates of the four heavy atoms $N^*, C_\alpha^*, C^*, O^*\in\mathbb{R}^3$ centered at $C_\alpha^*=\vec{0}$, assuming experimentally measured bond lengths and angles~\citep{bondangles}.
The main backbone atomic coordinates for a residue $i$ are given through
\begin{equation}
    [N_i, C_i, C_{\alpha,i}] = T_i\cdot[N^*, C^*, C^*_\alpha]\,,
\end{equation}
where $T_i\in\mathrm{SE}(3)$ is a member of the special Euclidean group, the set of valid translations and rotations in Euclidean space.
A backbone consists of $N$ frames $[T_1,\dots T_N]\in\mathrm{SE}(3)^N$, with the oxygen atom $O$ being reconstructed from an additional torsion angle $\psi\in\mathrm{SO}(2)$ around the $C_\alpha$ and $C$ bond.
Each frame is decomposed into $T_i=(r_i, x_i)$, where $x_i\in\mathbb{R}^3$ is the $C_\alpha$ translation and $r_i\in\mathrm{SO}(3)$ is a $3\times3$ rotation matrix which can be derived from relative atom positions with the Gram-Schmidt process.
A diffusion process over $\mathrm{SE}(3)^N$ can be constructed to achieve global SE(3) invariance by keeping the diffusion process centered at the origin.

We model the distribution over $\mathrm{SE}(3)^N$ through Riemannian score-based generative modeling, which aims to sample from a distribution supported on a Riemmanian manifold $\mathcal{M}$ by reversing a forward process that evolves from the data distribution $p_0$ towards an invariant density $p_T$ through
\begin{equation}
    d\mathbf{X}_t = -\tfrac12 \nabla U(\mathbf{X}_t)dt
    + d\mathbf{B}_{t,\mathcal{M}}\,,\quad \mathbf{X}_0\sim p_0\,,
\end{equation}
where $\mathbf{B}_{t,\mathcal{M}}$ is the Brownian motion on $\mathcal{M}$,  $U(x)$ is a continuously differentiable variable defining the invariant density $p_T\propto e^{-U(x)}$, $\nabla$ is the Riemannian gradient, and $t\in[0,T]$ is a continuous time variable. The  time-reversed process for $\mathbf{Y}_t=\mathbf{X}_{T-t}$ also satisfies a stochastic differential equation given by
\begin{equation}
    \label{eq:reverse}
    d\mathbf{Y}_t = \big[\tfrac12\nabla U(\mathbf{Y}_t) 
    + \nabla\log p_{T-t}(\mathbf{Y}_t)\big]dt + d\mathbf{B}_{t,\mathcal{M}}\,,\quad
    \mathbf{Y}_0 \sim p_T\,,
\end{equation}
where $p_t$ is the density of $\mathbf{X}_t$. 
The Riemannian gradients and Brownian motion depend on a choice of inner product on $\mathcal{M}$, which for SE(3) can simply be derived from the canonical inner products on SO(3) and $\mathbb{R}^3$.
The invariant density on SE(3) is chosen as $p_T \propto\mathcal{U}^{\mathrm{SO}(3)}(r)\,\mathcal{N}(x)$.

The Stein score $\nabla \log p_t$ itself is intractable and is therefore approximated with a score network $s_\theta$ which is trained with a denoising score matching loss given by
\begin{equation}
    \mathcal{L}_\mathrm{DSM}(\theta) = 
    \mathbb{E}\big[\lambda_t\|\nabla\log p_{t|0}(\mathbf{X}_t|\mathbf{X}_0)
    - s_\theta(t,\mathbf{X}_t)\|^2\big]\,,
\end{equation}
where $\lambda_t$ is a weighting schedule, $p_{t|0}$ is the density of $\mathbf{X}_t$ given $\mathbf{X}_0$, and the expectation is taken over $t$ and the distribution of $(\mathbf{X}_0,\mathbf{X}_t)$.
The loss on SE(3) is decomposed into its translation and rotation components as $\mathcal{L}_\mathrm{DSM} = \mathcal{L}^x_\mathrm{DSM} + \mathcal{L}^r_\mathrm{DSM}$.

To mitigate chain breaks or steric clashes and to learn the torsion angle $\psi$, two auxiliary losses are used. The first one is a direct mean squared error on the backbone positions $\mathcal{L}_{bb}$, while the second one is a local neighbourhood loss on pairwise atomic distances $\mathcal{L}_{2D}$. 
These losses are applied with a weight $w$ when sampling $t$ near 0, when fine-grained characteristics of the protein backbone emerge, such that the full training loss is expressed as
\begin{equation}
    \mathcal{L} = \mathcal{L}_{\mathrm{DSM}} 
    + w\,\Theta\big(\tfrac{T}{4}-t\big)\big(\mathcal{L}_{bb} + \mathcal{L}_{2D}\big)\,.
\end{equation}
The score network is based on the structure module of AlphaFold2~\citep{af2} and performs iterative updates over $L$ layers by combining spatial and sequence based attention modules using an Invariant Point Attention and a Transformer~\citep{transformer}, considering a fully connected graph structure.
As well as a denoised frame, the network also predicts the torsion angle $\psi$ for each residue, from which the positions of the backbone oxygen atoms can be reconstructed.

Sampling is achieved through an Euler-Maruyama discretisation of equation~(\ref{eq:reverse}) which is approximated with a geodesic random walk~\citep{grw}.
To avoid destabilisation of the backbone in the final sampling steps, trajectories are instead truncated at a time $\epsilon>0$.
For all numerical applications, we use identical parameters to the original {\framediff} model~\citep{framediff}.

We train this SE(3) diffusion model on synthetic antibody structural data, specifically targeting the variable domains whose CDR loops play a key role in defining the binding properties of the antibody.
Our dataset consists of 148,832 variable regions from the Observed Antibody Space (OAS)~\citep{oas1,oas2}, a database of antibody sequences, for which structures were predicted with {\abbtwo} ({\abbtwoshort})~\citep{abb2}, an antibody structure prediction model based on the structure module of AlphaFold-Multimer~\citep{afm}.

Our {\igdiff} model is obtained by fine-tuning the updated {\framediff} weights from July 2023 for 18 hours on 8 NVIDIA A10G GPUs, using an Adam optimizer~\citep{adam} with a learning rate of $10^{-4}$ and a batch size of 64. To encode the chain break between the heavy and light domains, we add a jump of 50 in the residue index across chains. We also include a heavy or light chain label as an additional node feature. The training data is obtained by clustering the concatenated CDR heavy and light chain sequences at 80\% similarity using CDHIT~\citep{cdhit}. For each epoch, we train on one representative from each cluster. Batches are constructed from equal length antibodies, where we complete a batch with previously seen antibodies if necessary.

\begin{figure}
    \centering
    \includegraphics[width=\textwidth]{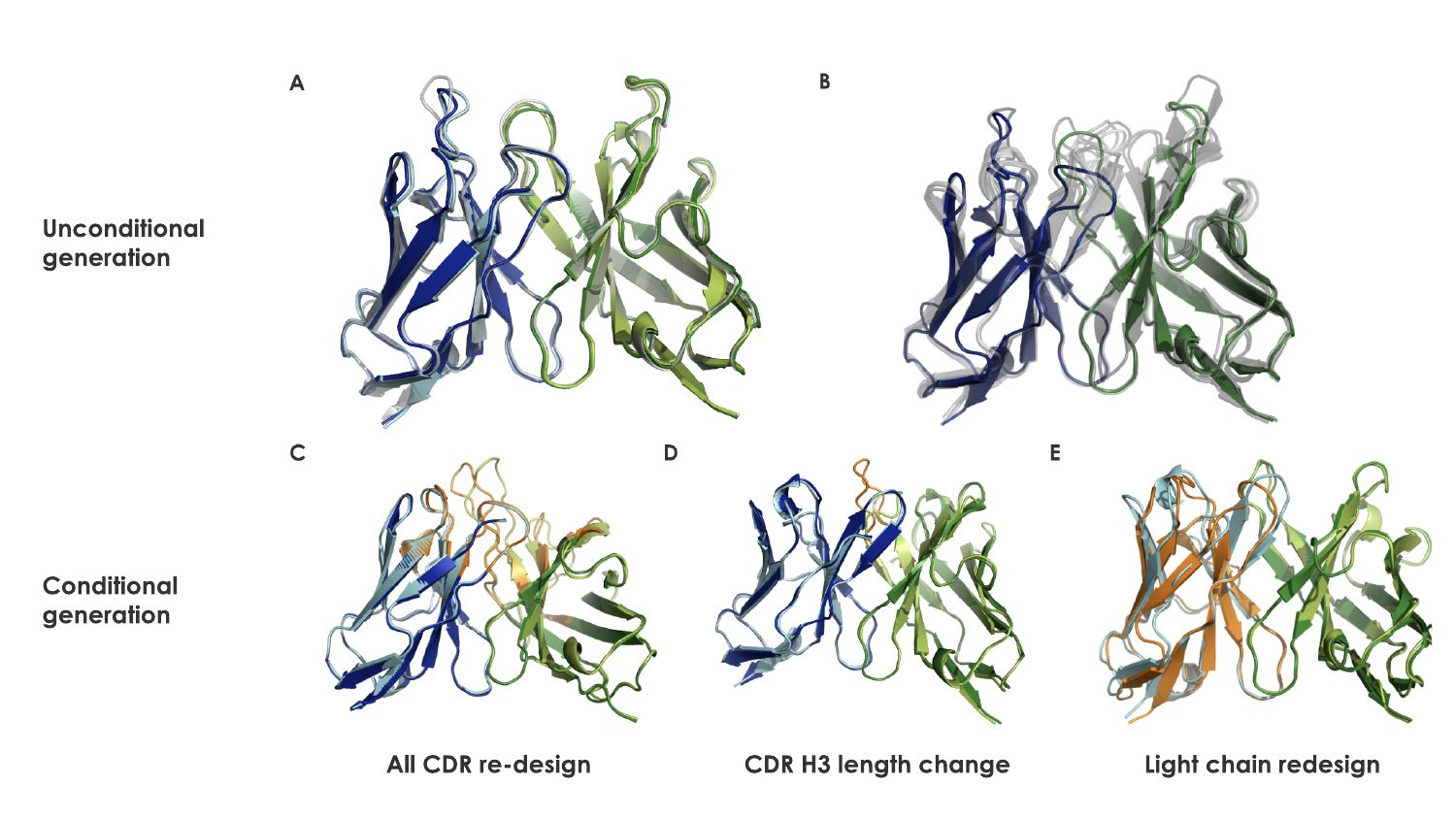}
    \caption{Examples of {\igdiff} generated antibody structures. Light chains are highlighted in blue, heavy chains in green. \textbf{(A-B)} Unconditionally generated antibodies. \textbf{(A)} {\igdiff} generated antibody (dark green/dark blue) compared to {\abbtwo} prediction on the lowest self-consistency RMSD sequence predicted by AbMPNN (light green/cyan) and the closest match in the training set by TM-score (grey). \textbf{(B)} Multiple antibodies generated by {\igdiff} with the same heavy and light chain length settings. \textbf{(C-E)} Conditionally generated antibodies, designed regions in orange, non-designed regions in dark blue/dark green, original input structure in cyan/light green. \textbf{(C)} Conditional generation of all CDR loops. \textbf{(D)} Conditional generation of CDR H3 with a different length compared to the input structure. \textbf{(E)} Conditional generation of the entire light chain.}
    \label{fig:pymol}
\end{figure}

In Figure~\ref{fig:pymol}, we show examples of unconditioned structures obtained with {\igdiff}, as well as conditioned samples for each of the design tasks considered in this study.
In order to determine the CDR regions, we use the IMGT numbering scheme~\citep{imgt} and North definition~\citep{north} throughout our analysis.

\FloatBarrier

\section{Generating \textit{de novo} antibody variable domains}

Let us now consider samples generated with {\igdiff} and study their properties.
We first analyse the ability of our {\igdiff} model to sample plausible, novel, and diverse antibodies, when generating the entire variable domain backbone of both heavy and light chains. 

Using our fine-tuned model, we start by sampling unconditioned paired heavy and light variable region structures. We obtain 100 unique combinations of chain lengths by sampling uniformly for the heavy and light chain independently between the 2.5 and 97.5 percentiles of the corresponding lengths found in the paired OAS. We then sample 8 structures for each of the 100 unique combinations for a total of 800 unique structures generated with {\igdiff}.

Sequences are predicted using the antibody-specific inverse folding model AbMPNN~\citep{abmpnn}, an adaptation of the general protein model ProteinMPNN~\citep{proteinmpnn} to antibodies. We sample 20 sequences for each generated structure with temperature $T=0.2$. To determine the best sequence for a given structure, we fold the sequence with {\abbtwoshort} and compare it to the {\igdiff} structure. The root mean squared deviation (RMSD) between the model predicted for a sequence and the initial generated structure is referred to as the self-consistency RMSD (scRMSD). For each generated sample, we keep the sequence that results in the lowest scRMSD.

For consistency and diversity metrics, we compare to a baseline of 800 structures drawn from the training set of {\abbtwoshort} predictions on paired OAS. For a fair comparison, we select 8 structures for each unique length combination of light and heavy chains in the {\igdiff} unconditioned dataset.  

\subsection{Consistency}

 \begin{figure}
     \centering
     \includegraphics[width=0.5\linewidth]{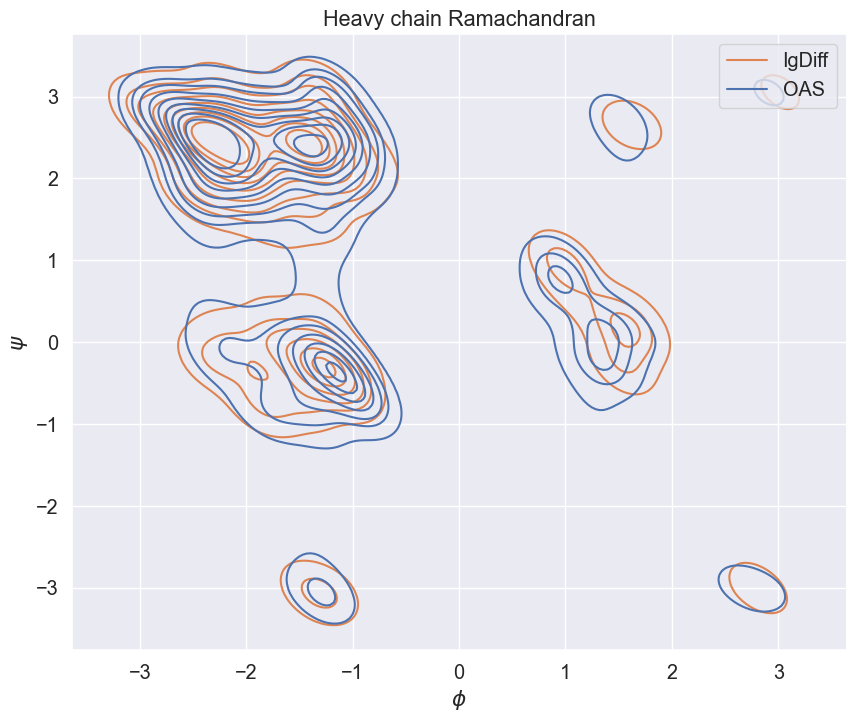}%
     \includegraphics[width=0.5\linewidth]{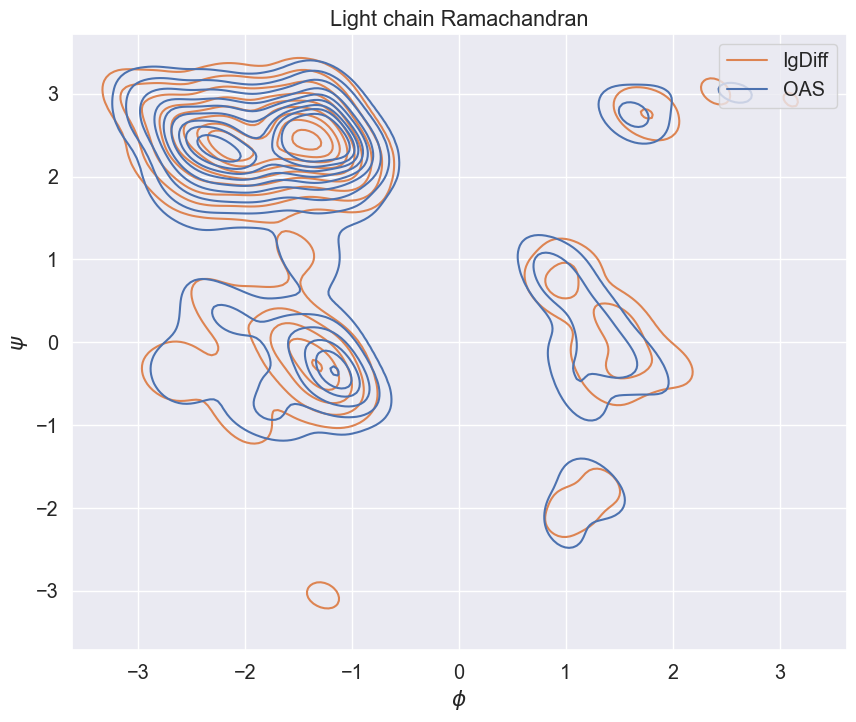}
     \caption{Left: Ramachandran plot of the dihedral angle distribution comparing the heavy chain residues from the predicted structures of OAS to {\igdiff}. Right: Same comparison for the light chain.}
     \label{fig:ramachandran}
 \end{figure}

We demonstrate that the {\igdiff} generated structures follow consistent structural distributions by comparing their backbone dihedral angle distributions in Ramachandran space to that of the OAS dataset. In Fig.~\ref{fig:ramachandran}, we show that the {\igdiff} $\phi$ and $\psi$ angle distribution closely matches that of the structures of naturally occurring antibody sequences from the paired OAS predicted with the modelling tool {\abbtwoshort}.

All generated antibodies have an scRMSD for the best predicted sequence that is below 2{\AA} across the entire antibody, while 88\% pass this threshold across each region independently, including all CDR loops. We also assess the consistency by considering the canonical clusters assumed by the CDR loops~\citep{chothia} and comparing those with the repredicted {\abbtwoshort} structures, finding that across each region, more than 91\% are assigned to the same canonical form.
Further studies of the scRMSD of generated antibodies and the distribution of canonical clusters in the CDR loops are given in Appendix~\ref{app:unconditioned}.

When {\abbtwoshort} outputs a prediction of a structure from a sequence, it first predicts an ensemble of four structures, and then selects the structure closest to the mean structure. This also allows it to provide a predicted error metric for each residue, computed from the mean deviation of the predicted residue position across the ensemble. When averaged across a set of residues this metric is  referred to as the root mean square predicted error or RMSPE. We use the RMSPE evaluated across the full variable domain as an additional check for consistency, for which a generated structure will pass if the RMSPE is less than the 90th percentile of the RMSPE on the {\abbtwoshort} testset. We find that 79.1\% of our generated structures pass this test.

\subsection{Novelty and diversity}
\begin{figure}
    \centering
    \includegraphics[width=0.265\textwidth]{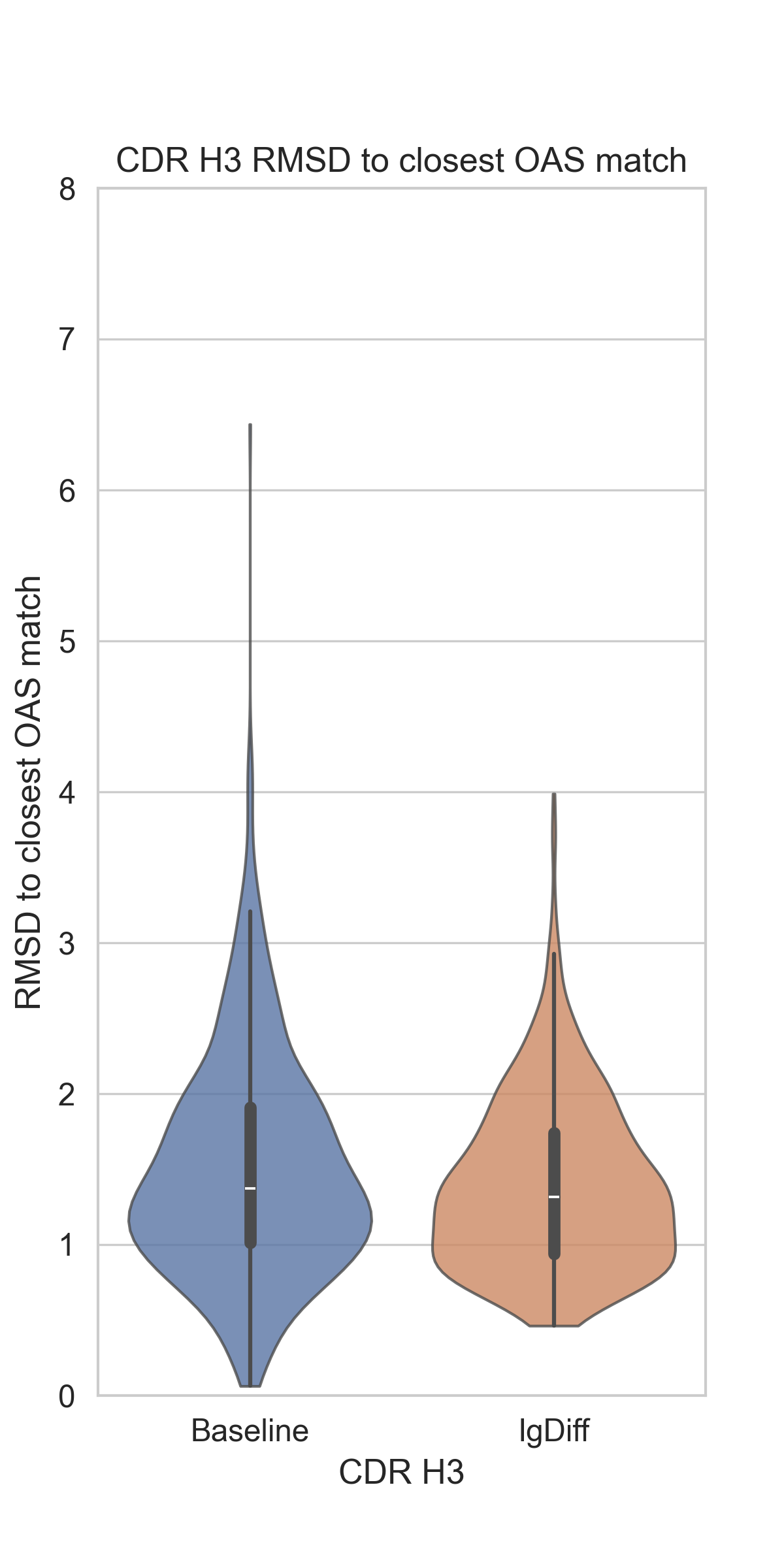}%
    \includegraphics[width=0.735\textwidth]{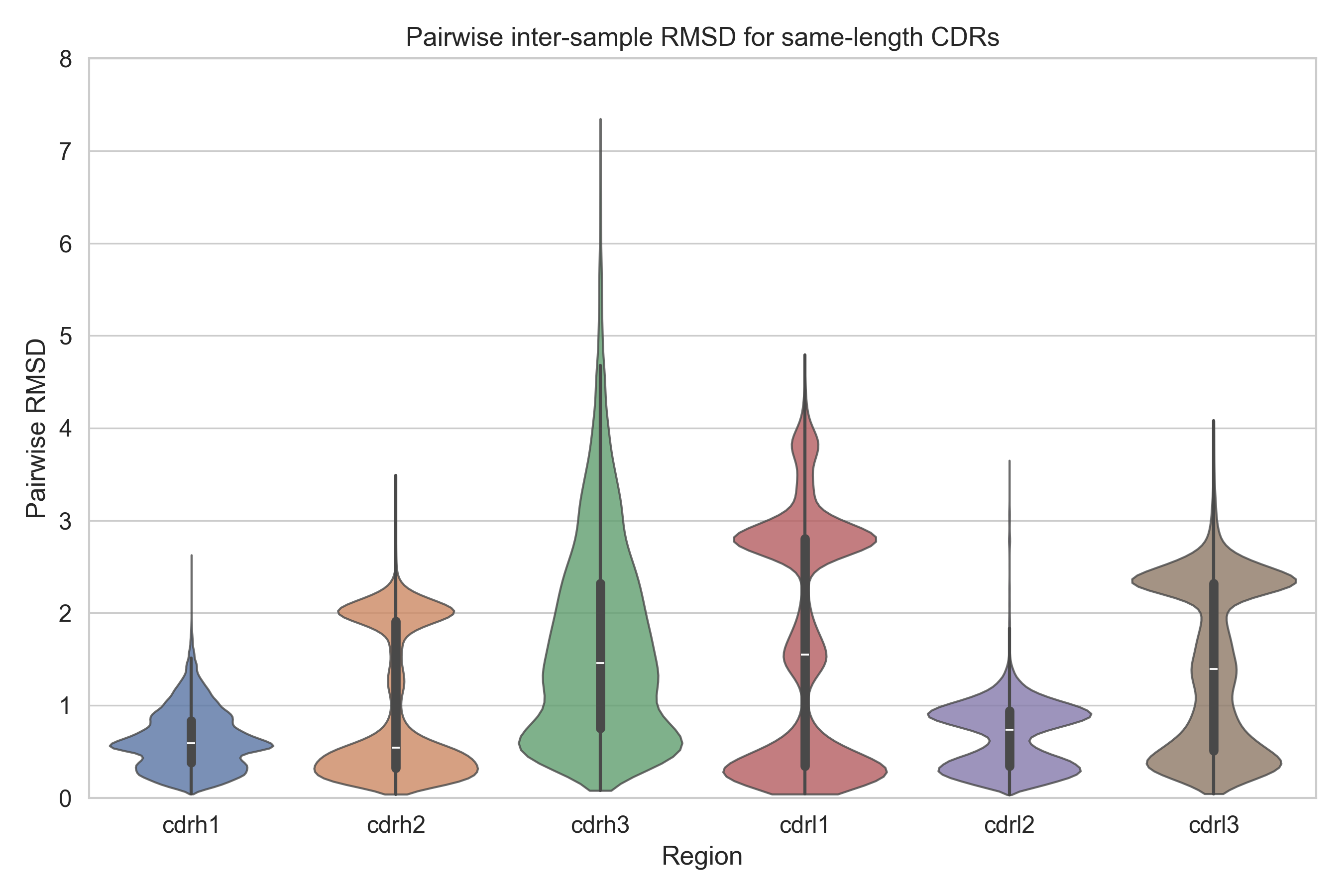}
    \caption{Left: Novelty of generated CDR H3 structures compared to random samples from the OAS dataset. For each structure, the RMSD shown is the CDR H3 RMSD to the closest match in the OAS dataset by TM-score with the same CDR H3 length. Right: Pairwise RMSD between {\igdiff} generated CDR loops of the same length. \textcolor{red}}
    \label{fig:novelty-diversity}
\end{figure}

We investigate the novelty of antibodies generated by {\igdiff} without conditioning by calculating the CDR H3 RMSD to the closest match by TM-score~\citep{tmscore} with identical CDR H3 loop length in the training dataset. This is shown in Fig.~\ref{fig:novelty-diversity} (left), comparing the distribution of CDR H3 RMSD of {\igdiff} generated antibodies to that of a randomly drawn set of antibodies from the training dataset with the same distribution of CDR H3 loop lengths. Generated antibodies have similar novelty as randomly drawn antibodies from the training dataset, with mean CDR H3 RMSD to the closest match of 1.39A $\pm$ 0.56 for {\igdiff} generated antibodies and 1.50A $\pm$ 0.74 for naturally observed antibodies, demonstrating that the model can produce novel H3 loops. 

We consider also the diversity of antibodies generated by {\igdiff}, an important consideration for antibody engineering. 
Taking CDRs with the same loop length, regardless of specified chain length, we calculate the pairwise RMSD between loops. This is shown in Fig.~\ref{fig:novelty-diversity} (right), where we can observe diversity of generated CDR loops across different samples. 
Interestingly, we observe a bimodal distribution of pairwise RMSDs for CDRs L1, L3 and H2, arising from two generated loops either assuming the same or different canonical conformations. 
Further studies of the distribution of loop lengths as a function of specified chain length and exploration of sequence diversity are discussed in Appendix~\ref{app:unconditioned}.

\subsection{Experimental validation}

In order to demonstrate that {\igdiff} generated antibodies are able to express and exhibit favorable designability characteristics, we selected 28 antibody sequences for experimental validation. Prior to selection we first excluded any of the sequences that had higher than the median CDRH3 scRMSD or higher than the median overall scRMSD. 
As detailed in Appendix~\ref{app:conditioned} all selected antibodies expressed and yielded sufficiently high concentration for downstream characterisation.

\section{Antibody design tasks}
Targeted design of specific regions of an antibody is of interest in a number of important applications. The engineering of the interface region, notably the CDR loops, can be of particular relevance for therapeutic applications~\citep{greiff,absci}. Furthermore, the pairing of an appropriate heavy or light chain to a given domain is of importance in antibody discovery and in shaping antibody repertoires~\citep{pairing}.

We thus consider the performance of {\igdiff} on conditional sampling for several inpainting design tasks of particular relevance to antibody engineering.
The tasks we consider are (i) the design of all CDR loops given a fixed heavy and light chain framework, (ii) the design of a complete light chain given a heavy chain, and (iii) the design of a CDRH3 loop with varying length given the remaining variable region, thus allowing for additional contact interactions from the longer loop.

We generate 10 structures per reference structure and CDR length. All starting point antibodies are taken from the {\abbtwoshort} test set. 
In order to perform inpainting, at each time $t$ during inference we replace each frame in the fixed region with the corresponding frame from the reference structure after applying the forward noising process to time $t$. 

In table~\ref{tab:conditioned_benchmark}, we show self-consistency metrics and compare our {\igdiff} model against {\rfdiff} on each design task.
None of the light chains designed by RFDiffusion can be parsed by Anarci~\citep{anarci} after inverse folding, while {\igdiff} always generates valid light chains, of which 93.3\% pass our confidence and scRMSD test. The CDRH3 length change task leads to 74\% of {\igdiff} generated structures that pass our combined test, compared with only 6\% of RFDiffusion designs.
For the task of designing all CDR loops, none of the RFDiffusion structures pass the scRMSD test, while 4\% of the {\igdiff} structures have scRMSD below 2{\AA} across all regions. Note here that this low success rate in the case of {\igdiff} is primarily driven by a high scRMSD in the CDRH1 loop, while the remaining regions have a success rate above 75\%. In contrast RFDiffusion also has poor self-consistency in CDRH1, CDRH3, CDRL1, and CDRL3. 
Further analysis of the scRMSD distribution and canonical clusters of designs, as well as details of the starting point antibodies used for each design task, are given in Appendix~\ref{app:conditioned}.

\begin{table}
\centering
\resizebox{\textwidth}{!}{
\begin{tabular}{l|ccc|ccc}
\toprule
& \multicolumn{3}{c|}{\igdiff} & \multicolumn{3}{c}{\rfdiff} \\ 
Task & scRMSD & Confidence & Combined & scRMSD & Confidence & Combined   \\
\midrule
Unconditioned & 0.88 & 0.79 & 0.76 & - & - & - \\
Design all CDRs & 0.04 &0.80 &0.02 & 0.00 & 0.54 & 0.00 \\
Change CDRH3 length & 0.76 & 0.91 & 0.74 & 0.08 & 0.69 & 0.06\\
Design light chain & 0.93 & 1 & 0.93 & - & - & - \\
\bottomrule
\end{tabular}
}
\vspace{2px}
\caption{Success rate of inpainted samples for each design task and three different tests, comparing {\igdiff} and RFDiffusion.  The scRMSD test requires all regions to have scRMSD < 2{\AA} independently. The confidence test requires the {\abbtwoshort} RMSPE averaged over all residues to be less than the 90th percentile of the same metric evaluated on the {\abbtwoshort} test set. The combined test metric reports the fraction of designs passing both previous tests. 
}
\label{tab:conditioned_benchmark}
\end{table}

\section{Conclusions}
In this article, we introduce a model for \textit{de novo} antibody generation, {\igdiff}.
This model is derived from the recent SE(3) diffusion framework {\framediff}, by fine-tuning on antibody variable domains.
The weights of our {\igdiff} model are made publicly available~\citep{zenodo}.

We show that our antibody backbone model is able to recapitulate the expected backbone dihedral distribution, and studied the validity of the sequences recovered from generated samples using an antibody-specific inverse folding model.
Studying the designability of the generated structures by comparing them with structure predictions based on the corresponding sequences, we found excellent agreement.
We probed our model for novelty by finding the closest match in the training data for each sampled structure and found it could generate structures distinct from those in the training set. 
We selected a number of designs for experimental validation, finding that all generated antibodies express with high yield.
We further found that {\igdiff} generates diverse antibodies, particularly in the CDR loops that are important determinants of binding affinity, making it well suited for antibody design tasks.

Finally, we considered examples of antibody engineering tasks, such as the redesign of CDR loops or the generation of a light chain paired to a specified heavy chain, and demonstrated the applicability of our approach in practical use cases.
We designed metrics to assess the quality of generated structures, and showed substantial improvement over existing state-of-the-art protein diffusion models.

Diffusion models trained on antibodies offer a promising approach to accelerate drug design through data-driven generative AI.
In this article, we provide a key step towards the generation of viable therapeutic antibodies through structure-based diffusion.
Conditioning the generation of samples to express desired properties, as well as to target specified antigens, will be a crucial elements towards facilitating their application in therapeutic development.

\section*{Acknowledgements}
We are grateful to Henry Kenlay, Claire Marks, Douglas Pires, Aleksandr Kovaltsuk and Newton Wahome for useful discussions.

\clearpage
\appendix
\section{Designability and novelty of unconditioned generation}
\label{app:unconditioned}

\begin{figure}[ht]
    \centering
    \includegraphics[width=0.5\textwidth]{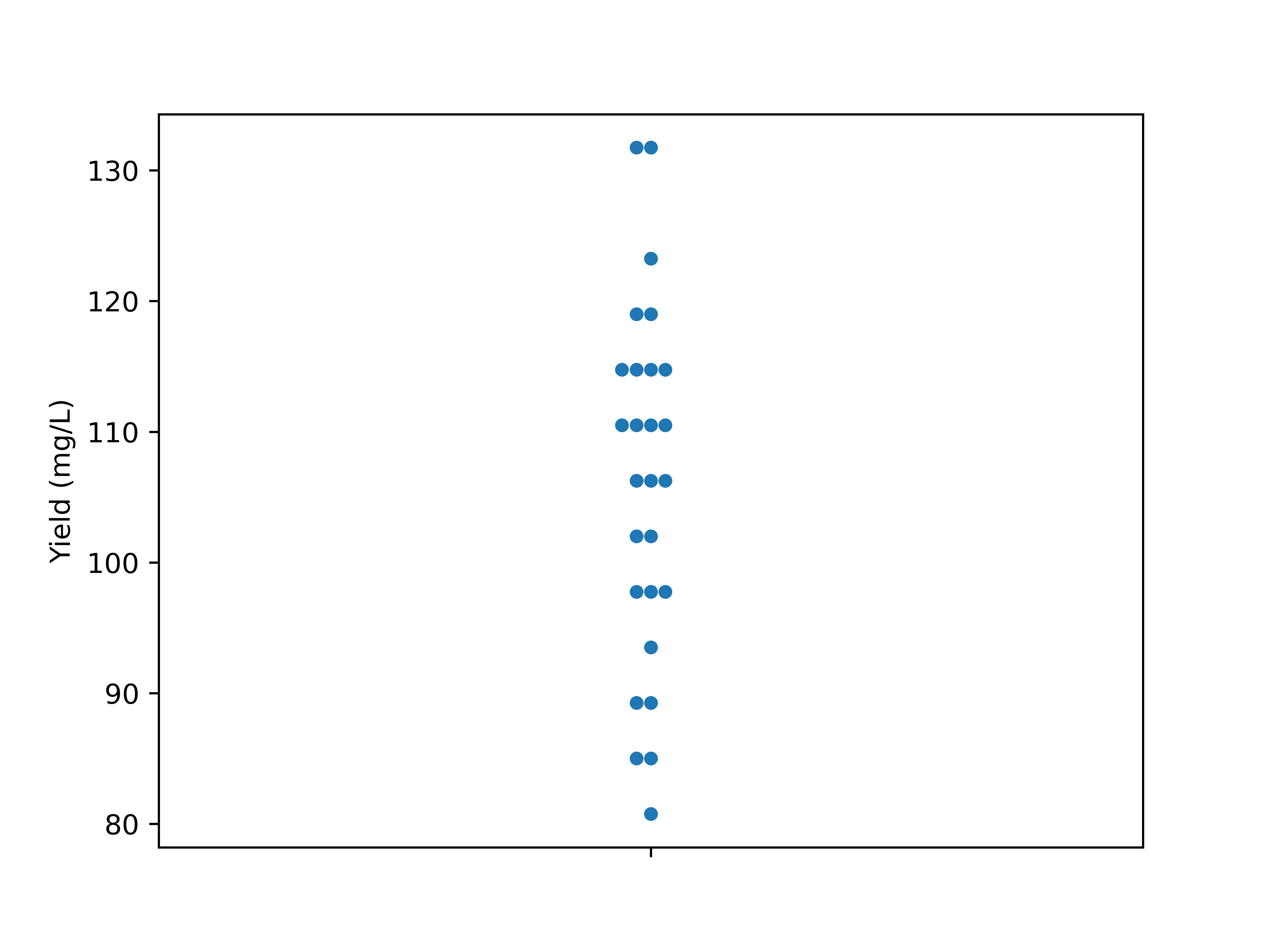}
    \caption{
    Expression yields (mg/L) of a validation set of {\igdiff} generated antibodies. 
    }
    \label{fig:yield}
\end{figure}

In Figure~\ref{fig:yield}, we show the expression yields for 28 selected antibody sequences predicted with AbMPNN from full unconditioned variable structures produced with {\igdiff}.

\begin{figure}[ht]
    \centering
    \includegraphics[width=0.9\textwidth]{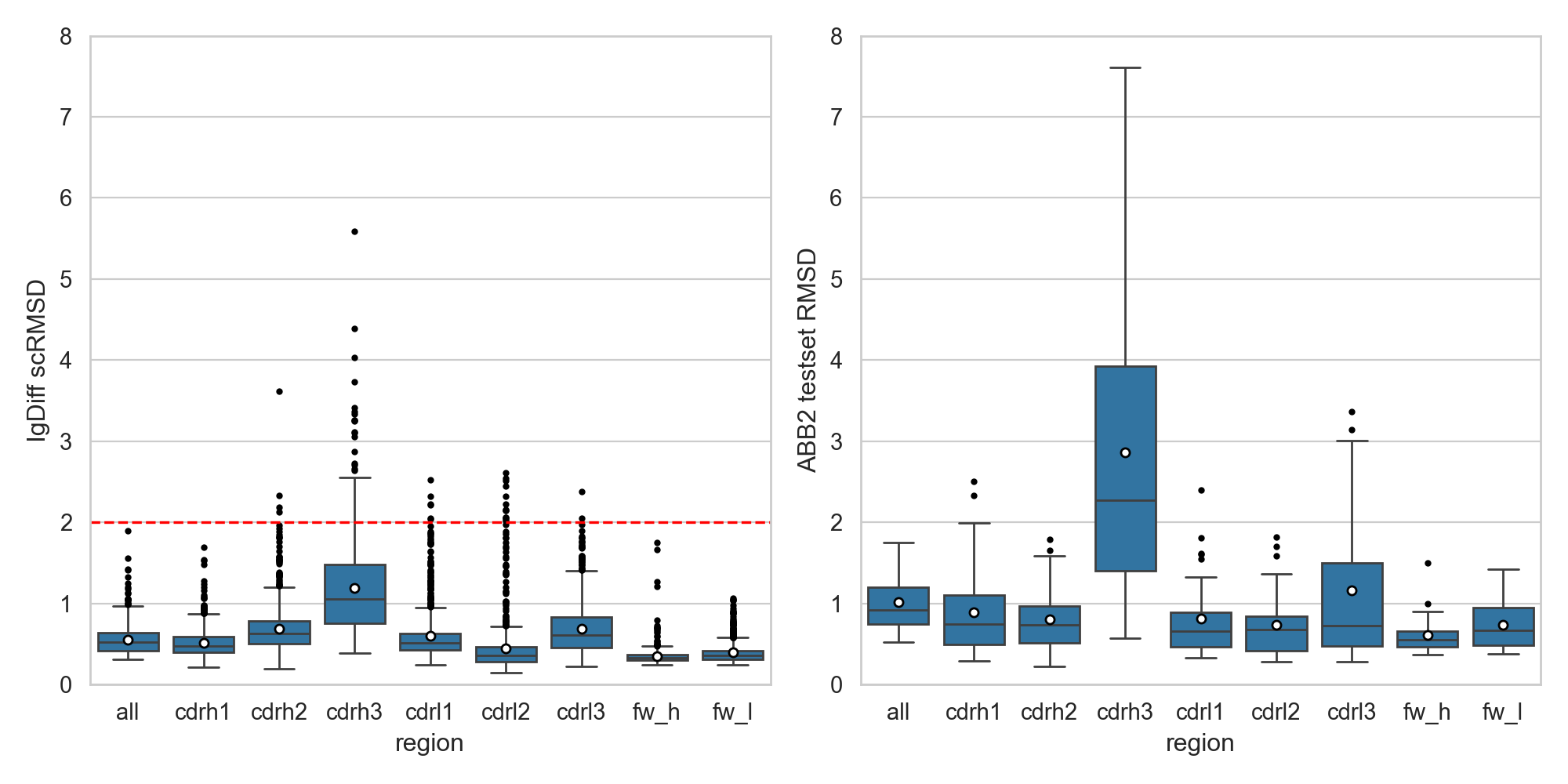}
    \caption{scRMSD of {\igdiff} structures against {\abbtwoshort} re-predictions, compared to inherent {\abbtwoshort} RMSD. \textbf{(Left)} {\igdiff} scRMSD to the {\abbtwoshort} predicted structure for the AbMPNN sequence producing the lowest overall scRMSD to the input {\igdiff} structure. The red line indicates the 2\AA~cutoff used to designate a designable structure in {\framediff}~\citep{framediff}. \textbf{(Right)} RMSD of {\abbtwoshort} predictions on the {\abbtwoshort} testset against the ground truth crystal structures. Mean {\igdiff} scRMSD on each region are lower than the corresponding {\abbtwoshort} test set RMSD.}
    \label{fig:scrmsd}
\end{figure}

We assess the scRMSD of {\igdiff} generated antibody structures to corresponding {\abbtwoshort} predicted structures.
To this end, we predict 20 amino acid sequences for each {\igdiff} generated backbone structure using AbMPNN and use {\abbtwoshort} to predict the structure those sequences are likely to assume, reporting the scRMSDs for the AbMPNN prediction that achieves the lowest overall scRMSD across the 20 predictions for each {\igdiff} output, as shown in Fig.~\ref{fig:scrmsd}. 
A lower scRMSD in this scenario indicates that the {\igdiff} generated antibody structures represent realistic and designable antibody structures. 
We show that the structures are designable in that the scRMSD in all loops are typically lower than the 2 \AA~cutoff used to define designability in the original {\framediff} paper~\citep{framediff}. 
All {\igdiff} generated antibodies pass the 2\AA~threshold across the entire antibody and 88\% of {\igdiff} generated antibodies also pass for every sub-region assessed independently in Fig.~\ref{fig:scrmsd}, including all CDR loops. 

We further assessed self-consistency through the canonical clusters assumed by the non-H3 CDR loops in both {\igdiff} generated structures and {\abbtwoshort} repredicted structures (generated as described for the scRMSD calculation). To this end, each loop is assigned to the closest PyIGClassify cluster centre by RMSD~\cite{adolf2015pyigclassify}. If a loop has an unusual length with no clusters or has $\text{RMSD} > 1.5${\AA} to the closest cluster centre then it is denoted as unclassified.

\begin{figure}
    \centering
    \includegraphics[width=1.0\textwidth]{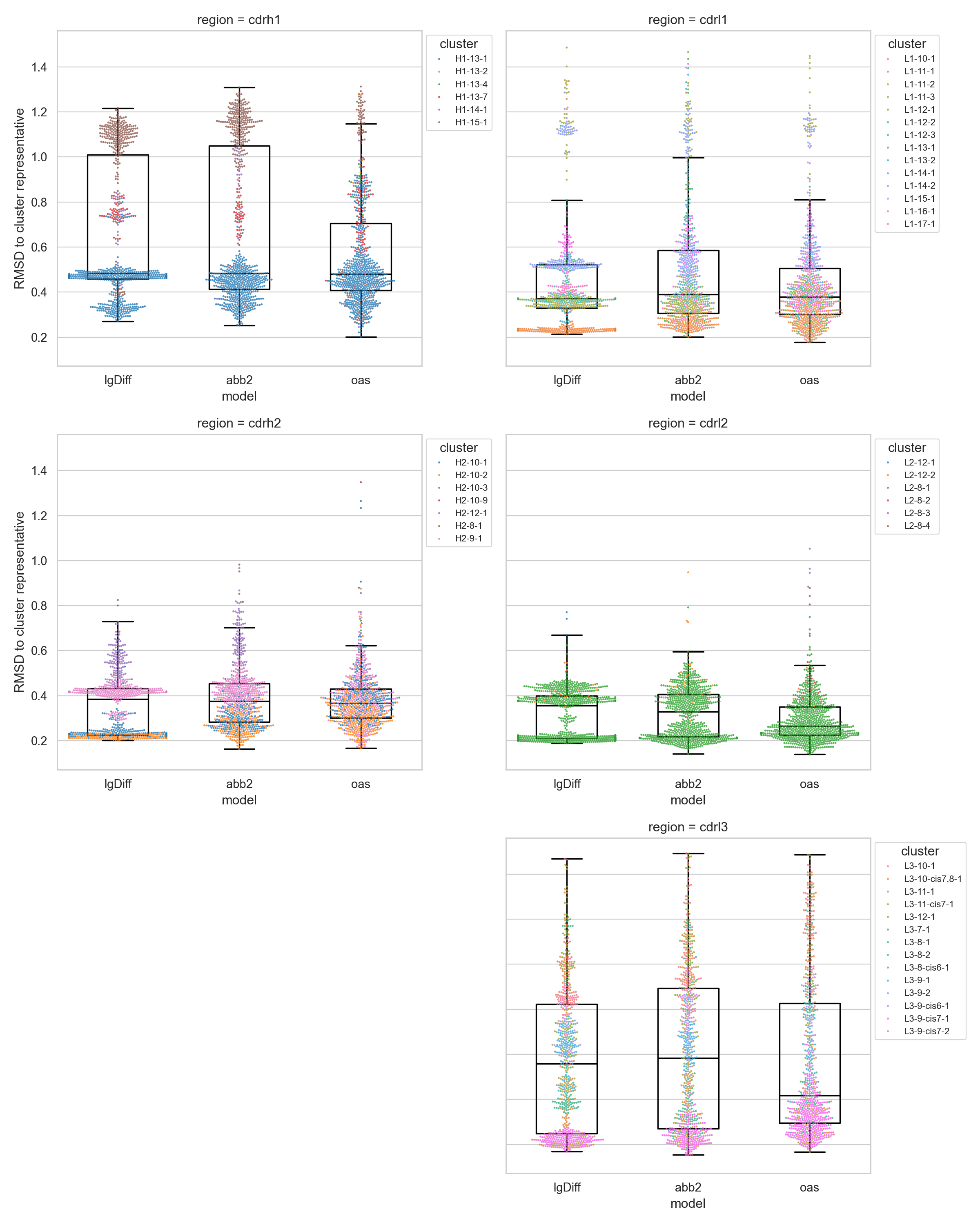}
    \caption{RMSD to the central cluster representative in the PyIGClassify dataset for all classified canonical loops. Each panel refers to a different CDR loop. For each panel we show the result for the generated {\igdiff} structures on the left, the results for the {\abbtwoshort} structures predicted on the {\igdiff} sequences in the centre, and a baseline of paired OAS structure predictions on the right. In each plot the individual datapoints that make up a box-plot are shown as a swarm-plot with cluster class indicated by the point hue.}
    \label{fig:cluster-rmsd}
\end{figure}

For non-H3 loops, {\igdiff} structures and {\abbtwoshort} repredicted structures were assigned the same canonical class in between 93\% (CDR L1) and 98\% (CDR H1) of samples  for the less variable CDRs 1 and 2 and in 85\% of samples for the more variable CDR L3. Up to 8\% (CDRH2) of generated designs fell outside of the known canonical class for that loop both in the {\igdiff} generated structure and in the {\abbtwoshort} repredicted structure (see Table~\ref{tab:unconditional_clusters}). We further show the RMSD to the assigned cluster centre for each loop generated by {\igdiff} and {\abbtwoshort} respectively, as well as for a baseline of OAS paired {\abbtwoshort} predicted structures (see Fig.~\ref{fig:cluster-rmsd}). We demonstrate that RMSD to the closest cluster centre depends heavily on the canonical class of the loop and that {\igdiff} generated antibodies follow a similar distribution of distances to the closest cluster centre as the {\abbtwoshort} repredicted structures and a baseline set of OAS derived, {\abbtwoshort} predicted structures. 
\begin{table}
\centering
\begin{tabular}{lccc}
\toprule
\multirow{2}{*}{Region} & 
\multirow{2}{*}{\begin{tabular}[c]{@{}l@{}}Matching \\ clusters\end{tabular}} & 
\multirow{2}{*}{\begin{tabular}[c]{@{}l@{}}Mismatching \\ clusters\end{tabular}} & 
\multirow{2}{*}{\begin{tabular}[c]{@{}l@{}}Both \\ unclassified\end{tabular}} \\ \\
\midrule
CDRH1 &  0.98 & 0.01 & 0.01 \\
CDRH2 &  0.92 & 0.00 & 0.08 \\
CDRL1 &  0.93 & 0.03 & 0.04 \\
CDRL2 &  0.95 & 0.01 & 0.04 \\
CDRL3 &  0.85 & 0.09 & 0.06 \\
\bottomrule\\
\end{tabular}
\caption{Canonical cluster analysis of antibody structures generated unconditionally using {\igdiff} and the {\abbtwoshort} repredicted structures.}
\label{tab:unconditional_clusters}
\end{table}

\begin{figure}
    \centering
    \includegraphics[width=0.5\textwidth]{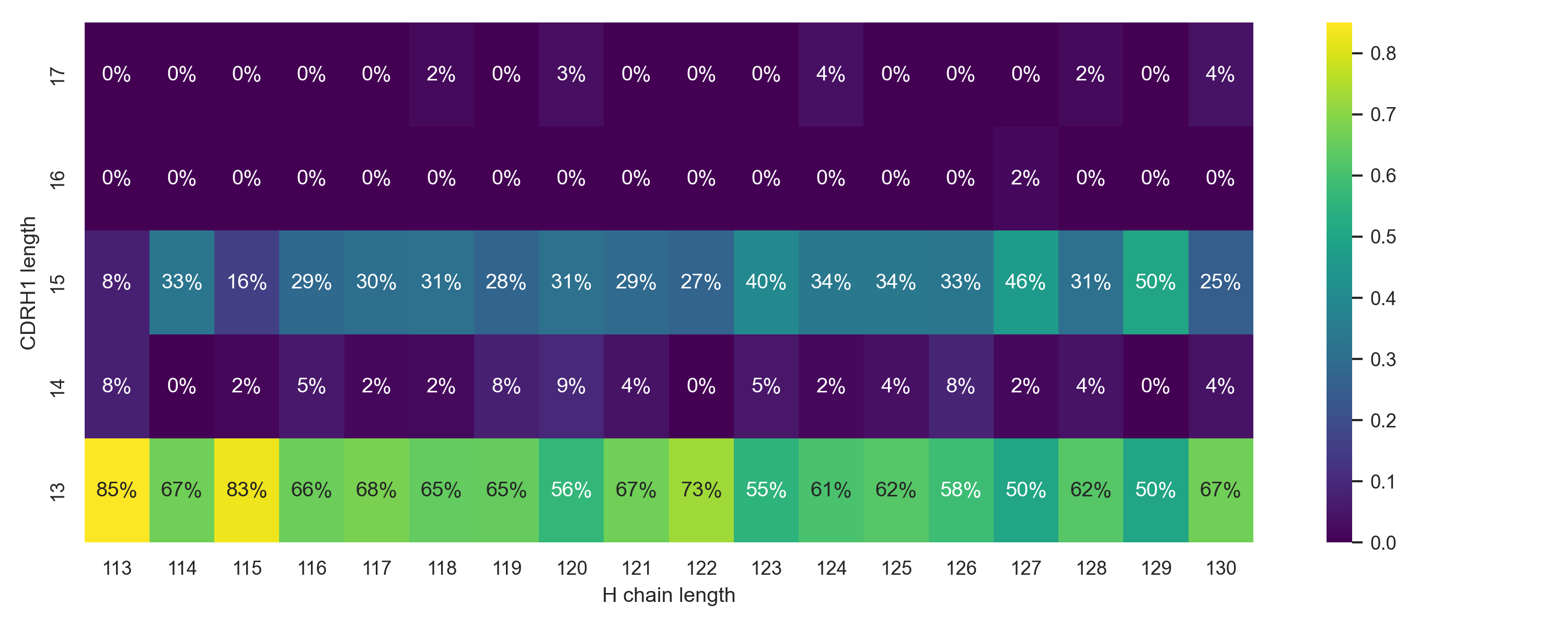}%
    \includegraphics[width=0.5\textwidth]{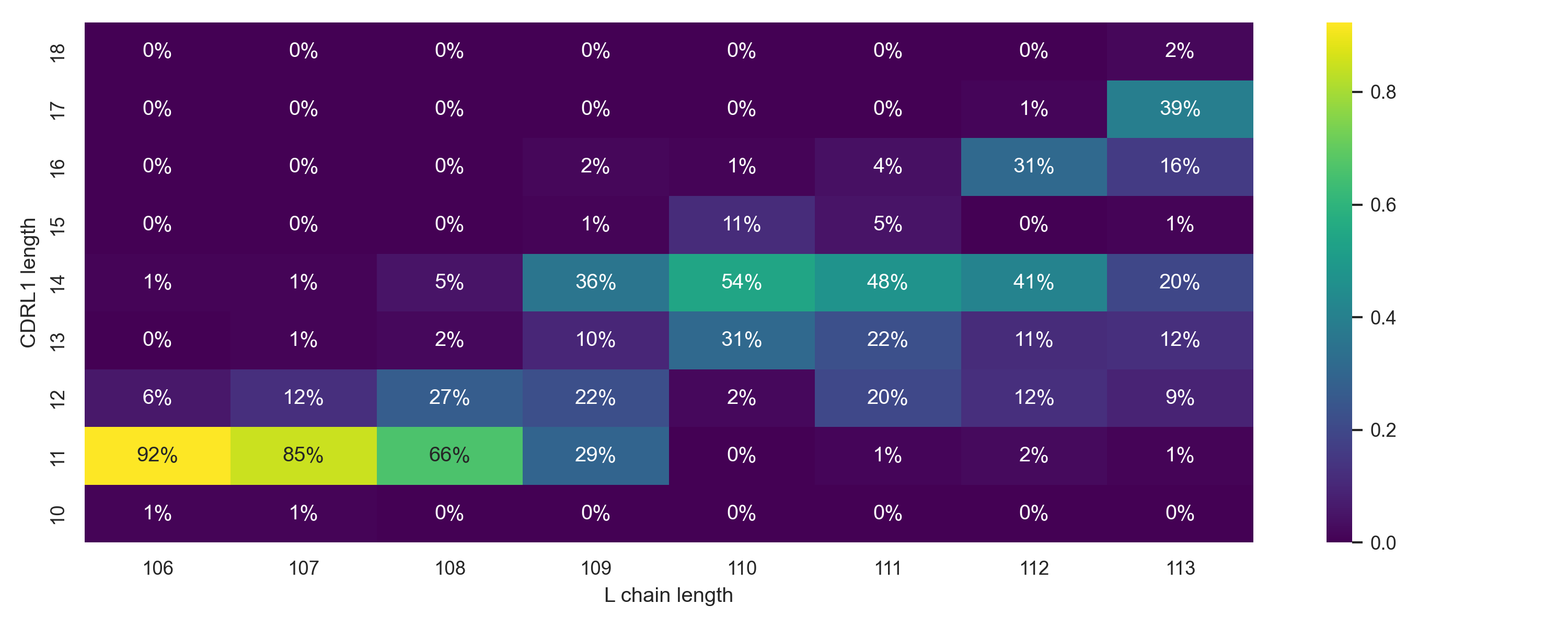}
    \includegraphics[width=0.5\textwidth]{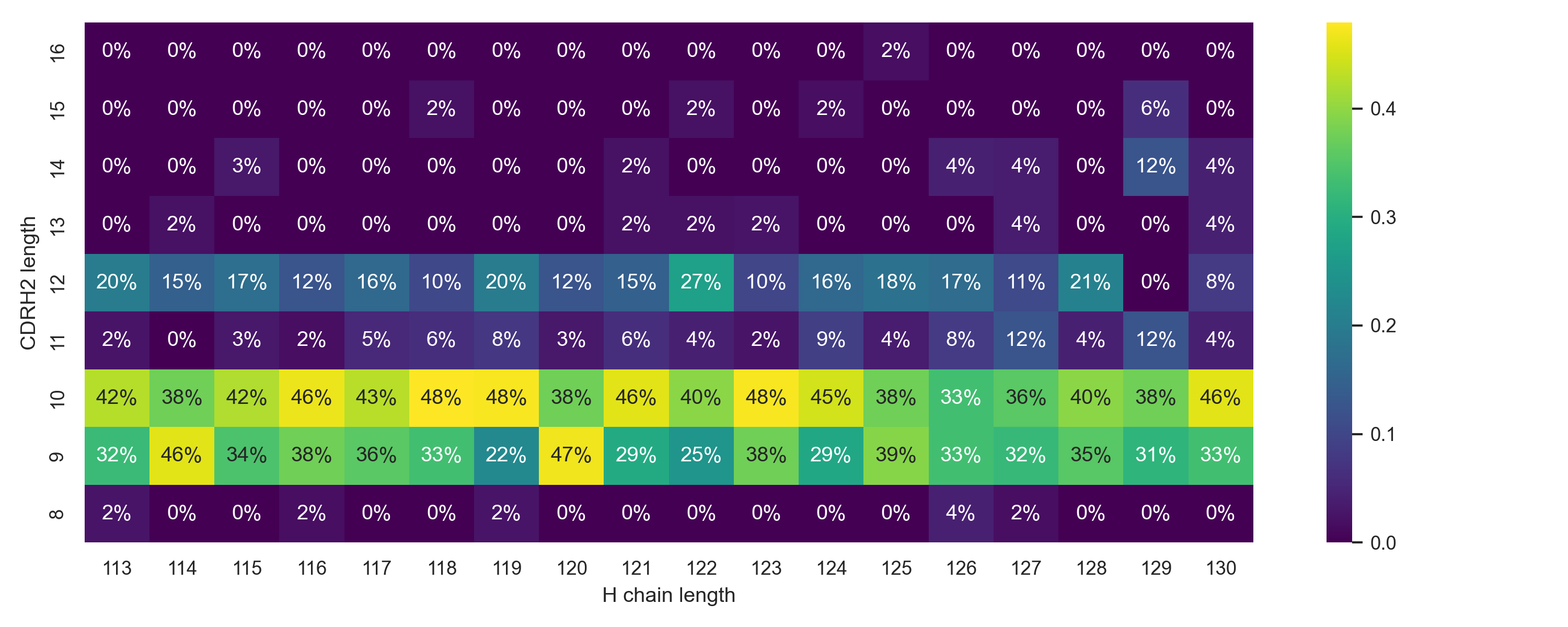}%
    \includegraphics[width=0.5\textwidth]{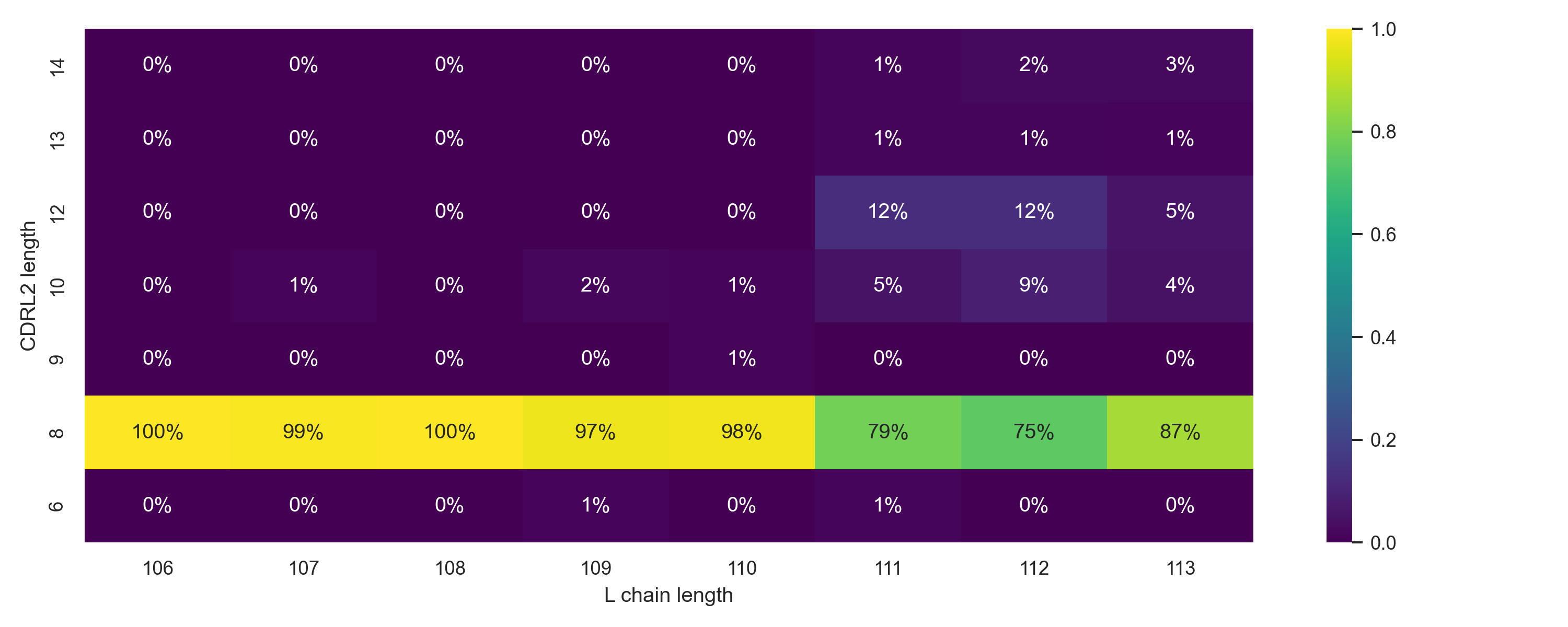}
    \includegraphics[width=0.5\textwidth]{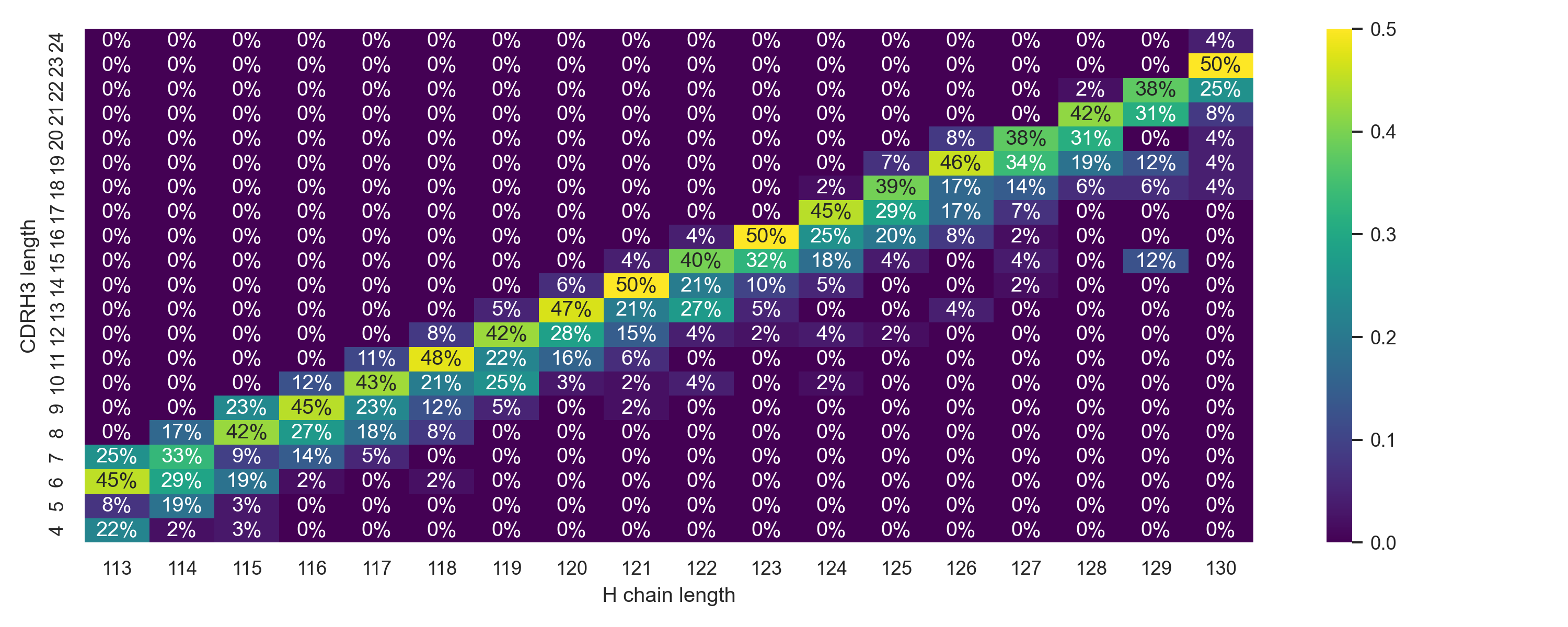}%
    \includegraphics[width=0.5\textwidth]{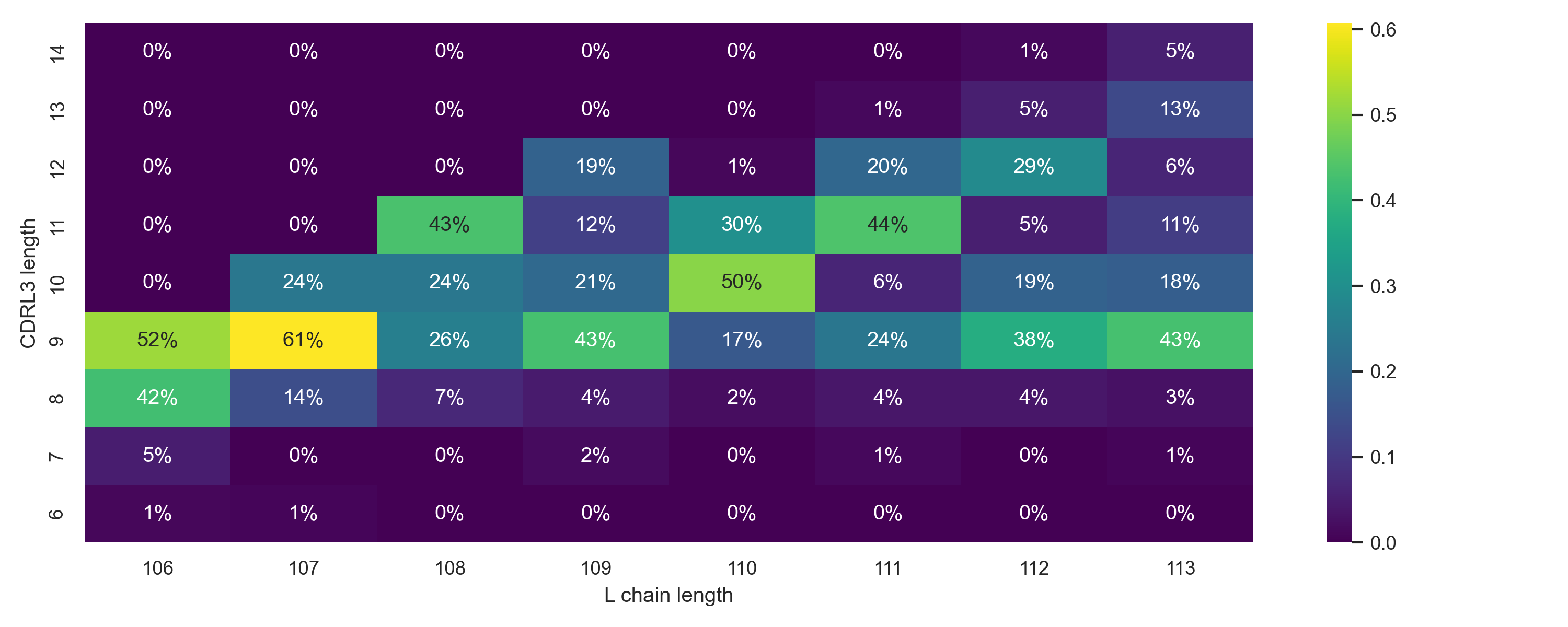}
    \caption{Distribution of CDR loop lengths depending on the generated heavy or light chain length.}
    \label{fig:generated-cdr-lengths}
\end{figure}

For heavy or light chains generated with the same length specification, we investigate the distribution of loop length for each CDR, as CDRs with different lengths can necessarily be considered diverse. This is shown in Fig.~\ref{fig:generated-cdr-lengths}. 
We observe that {\igdiff} generates diverse CDR loop lengths largely independent of the pre-specified chain length. 
For the light chain, diversity of CDR loop lengths increases with increasing chain length, while for the heavy chain, the diversity of generated CDR loops remains broadly constant across chain lengths (with the exception of CDR H1 at short chain lengths). 
We further observe that {\igdiff} drives chain length of the heavy chain primarily via CDR H3 length and light chain length via CDR L1 and L3 length, mirroring the natural distributions of CDR loop lengths.

\begin{figure}
    \centering
    \includegraphics[width=0.8\textwidth]{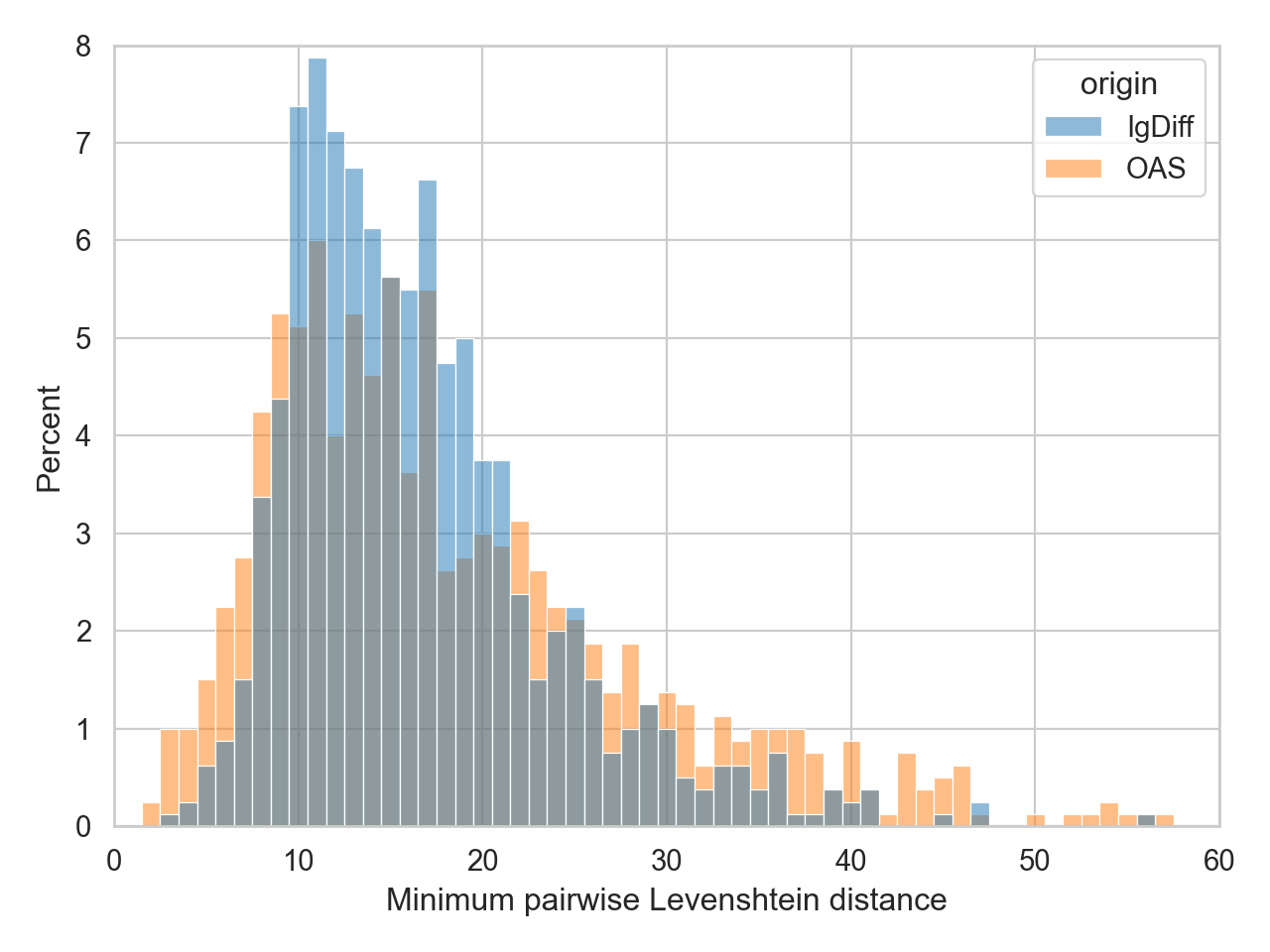}
    \caption{Histogram of the minimum pairwise Levenshtein distance between 800 unconditioned {\igdiff} generated structures (blue) and 800 paired OAS sequences (orange).}
    \label{fig:min-edit}
\end{figure}

To explore sequence diversity, we find the minimum pairwise Levenshtein distance (also known as edit distance) between the sequence of a generated {\igdiff} structure and the sequences in the rest of the generated structures. 
The resulting distribution of pairwise distances is shown in Fig.~\ref{fig:min-edit}. Each generated antibody has at least 3 edits between the closest sequence, and at most 56 edits. The median number of edits to the closest sequence is 15. For a comparison we also show the minimum pairwise edit distances between 800 example predicted structures taken from the paired OAS dataset with the same chain lengths as the {\igdiff} unconditioned dataset. Both distributions are similar, with {\igdiff} producing fewer antibodies with very low and very high minimum edit distances and more antibodies with medium edit distances than the OAS baseline. 

\clearpage

\section{Benchmarking of design tasks}
\label{app:conditioned}

The starting point antibodies used in each design task are given in Table~\ref{tab:design_tasks}. In this appendix, we consider also the task of redesigning the CDRL3 loop with varying lengths, to study the recapitulation of the canonical clusters of designs with modified loop lengths. This task is left out of the main text for conciseness but achieves 100\% success on the scRMSD test for {\igdiff} compared to 85\% for {\rfdiff}, with both models obtaining 100\% success on the confidence test.

\begin{table}[!ht]
    \centering
    \begin{tabular}{lllc}
        \toprule
        Design task  & Subtasks & Reference antibody & Total samples\\ \midrule
        CDRH3 length change & H3 lengths 10-19 & 7seg & 100 \\
        CDRL3 length change & L3 lengths 8-11 & 7sem & 40\\ 
        Design all CDRs & Different antibodies & 7ps6, 7q4q, 7rp2, 7ttm, 7u8c & 50\\ 
        Design light chain & Different heavy chains & 7qf0, 7rxl, 7zf6 & 30 \\ 
        \bottomrule
        \\
    \end{tabular}
    \caption{Conditional design tasks. We sample 10 structures for each inpainting subtask. The CDRL3 length change design task is limited to this Appendix.}
    \label{tab:design_tasks}
\end{table}

\begin{figure}[ht]
    \centering
    \includegraphics[width=\textwidth]{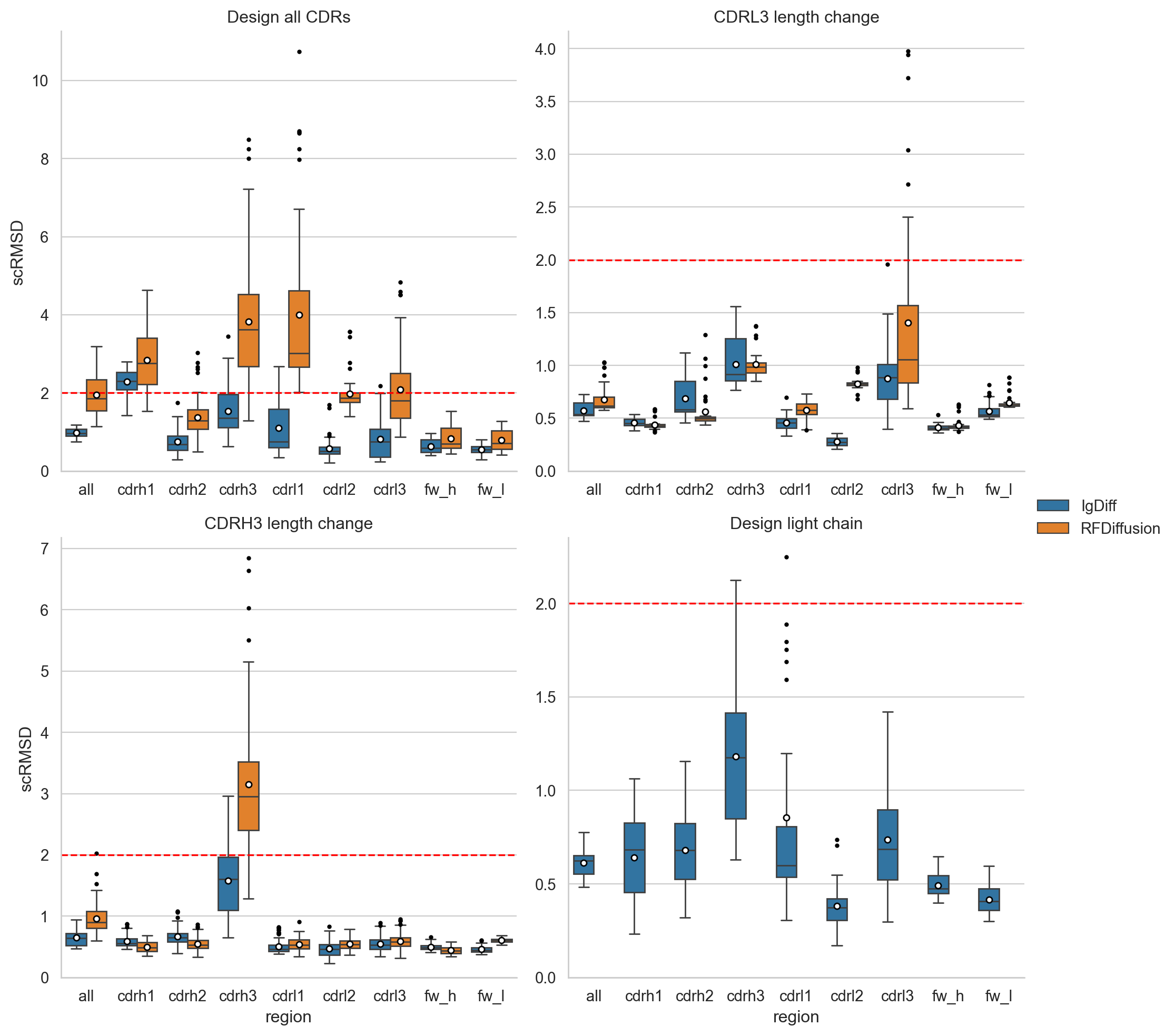}
    \caption{scRMSD of antibodies generated conditionally using {\igdiff} and {\rfdiff} with different design tasks. The red line indicates the 2{\AA}  cutoff used to designate designable structures.}
    \label{fig:scRMSD-conditional}
\end{figure}

\begin{figure}
    \centering
    \includegraphics[width=0.85\textwidth]{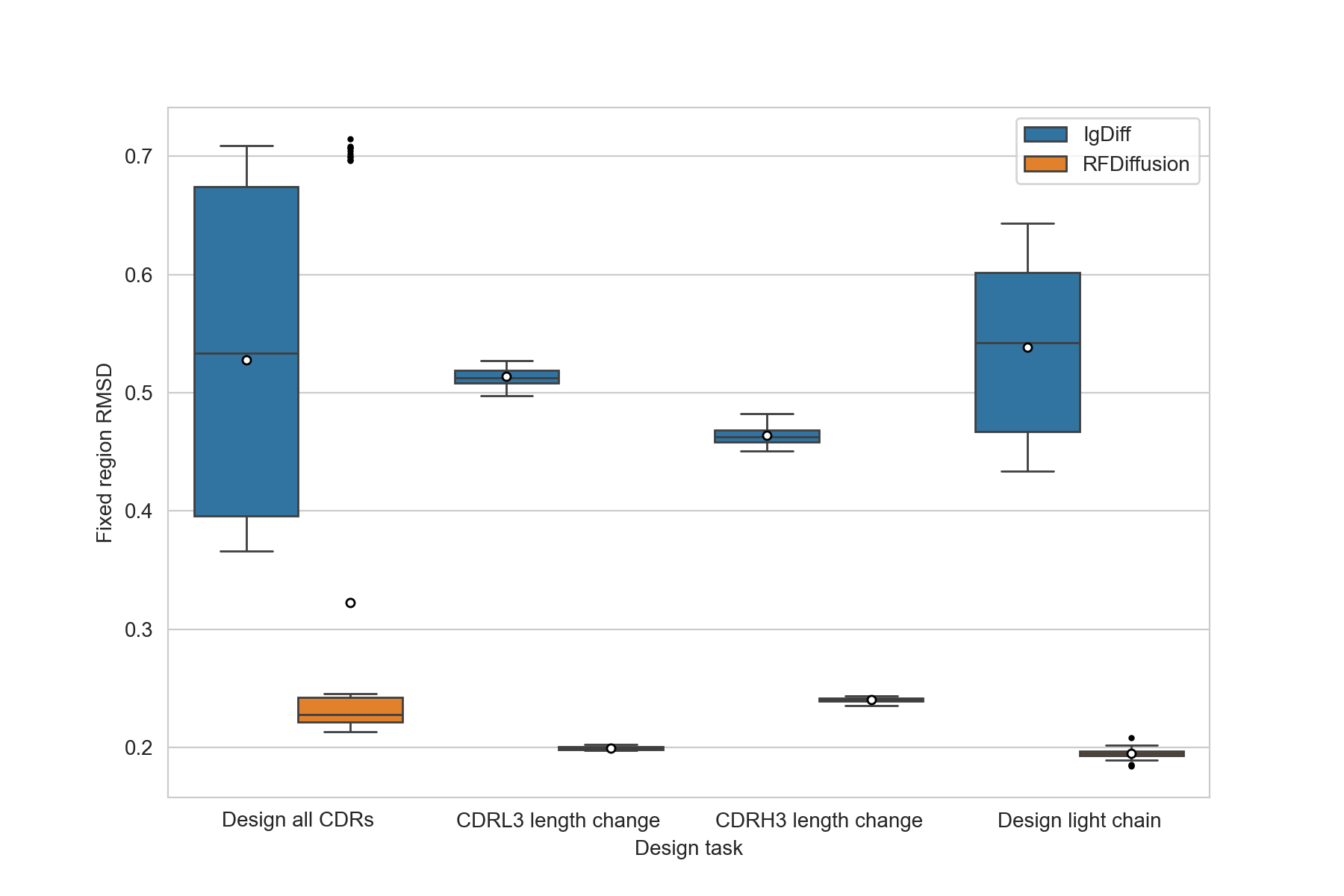}
    \caption{RMSD of the fixed regions during conditional design tasks using {\igdiff} and {\rfdiff}. For each design task, the RMSD is evaluated on the fixed region and between the generated design and the reference antibody}
    \label{fig:fixed-region-rmsd}
\end{figure}

In Figure~\ref{fig:scRMSD-conditional}, we show the scRMSD of each region across the different design tasks. We compare the performance of {\igdiff} to the baseline of {\rfdiff}. We see that in all design tasks {\igdiff} has a lower mean and median scRMSD in the region that is being inpainted. The most challenging task for {\igdiff} is to design all of the CDRs. Interestingly, the region that {\igdiff} most struggles to model in this conditional task appears to be CDRH1, although this region was well modelled in unconditional generation. For all other tasks, we show that almost all designed antibodies retain a scRMSD below 2{\AA} across all regions, indicating that the conditioned antibodies retain the designability properties of the unconditioned model. Figure~\ref{fig:fixed-region-rmsd} shows the RMSD with the input structure across the fixed regions of the design task. This is below 1{\AA} for all generated structures, demonstrating that the motifs are well preserved during the inpainting procedure.

In Table~\ref{tab:conditional_clusters}, we give the fraction of matching canonical clusters with repredicted {\abbtwoshort} structures for each design task. Here we note good agreement across all CDR loops except CDR H1. Notably {\igdiff} outperforms {\rfdiff} on the fraction of matching clusters on all tasks.

\begin{table}[ht]
\centering
\resizebox{\textwidth}{!}{
\begin{tabular}{llcccccc}
\toprule
 & & \multicolumn{3}{c}{\igdiff} & \multicolumn{3}{c}{\rfdiff} \\
\multirow{2}{*}{Task}& 
\multirow{2}{*}{Region} & 
\multirow{2}{*}{\begin{tabular}[c]{@{}l@{}}Matching \\ clusters\end{tabular}} & 
\multirow{2}{*}{\begin{tabular}[c]{@{}l@{}}Mismatching \\ clusters\end{tabular}} & 
\multirow{2}{*}{\begin{tabular}[c]{@{}l@{}}Both \\ unclassified\end{tabular}} &  
\multirow{2}{*}{\begin{tabular}[c]{@{}l@{}}Matching \\ clusters\end{tabular}} & 
\multirow{2}{*}{\begin{tabular}[c]{@{}l@{}}Mismatching \\ clusters\end{tabular}} & 
\multirow{2}{*}{\begin{tabular}[c]{@{}l@{}}Both \\ unclassified\end{tabular}}
\\ 
\\
\midrule
Design all CDRs & CDRH1&  0.02 & 0.96 & 0.02 & 0.06 & 0.84 & 0.10 \\
Design all CDRs & CDRH2&  1.00 & 0.00 & 0.00 & 1.00 & 0.00 & 0.00 \\
Design all CDRs & CDRL1&  0.70 & 0.24 & 0.06 & 0.02 & 0.74 & 0.24 \\
Design all CDRs & CDRL2&  0.98 & 0.02 & 0.00 & 0.02 & 0.98 & 0.00 \\
Design all CDRs & CDRL3&  0.90 & 0.10 & 0.00 & 0.46 & 0.54 & 0.00 \\
\midrule
CDRL3 length change & CDRL3&  0.70 & 0.28 & 0.03 & 0.68 & 0.33 & 0.00 \\
\midrule
Design light chain & CDRL1&  0.80 & 0.17 & 0.03 & -- & -- & -- \\
Design light chain & CDRL2&  1.00 & 0.00 & 0.00 & -- & -- & -- \\
Design light chain & CDRL3&  0.87 & 0.07 & 0.07 & -- & -- & -- \\
\bottomrule\\
\end{tabular}
}
\caption{Fraction of matching and mismatching canonical clusters between the antibody structures generated conditionally using {\igdiff} and the {\abbtwoshort} repredicted structures. Note that {\rfdiff} did not produce antibody light chains in the light chain design task.}
\label{tab:conditional_clusters}
\end{table}

\clearpage

\bibliography{references}
\bibliographystyle{plainnat}


\end{document}